\documentclass{aastex631}
\usepackage{CJK}
\usepackage{booktabs}
\usepackage{graphbox}
\usepackage{amsmath}
\usepackage{soul}

\begin{document}

\begin{CJK*}{UTF8}{gbsn}

\title{On the Determination of Stellar Mass and  Binary Fraction of Open Clusters within $500\,\rm pc$ from the Sun}

\correspondingauthor{Jing Zhong}
\email{jzhong@shao.ac.cn}

\author[0000-0002-4986-2408]{Yueyue Jiang (蒋悦悦)}
\affiliation{Key Laboratory for Research in Galaxies and Cosmology, Shanghai Astronomical Observatory, Chinese Academy of Sciences, 80 Nandan Road, Shanghai 200030, China}
\affiliation{School of Astronomy and Space Science, University of Chinese Academy of Sciences, No. 19A, Yuquan Road, Beijing 100049, China}

\author[0000-0001-5245-0335]{Jing Zhong (钟靖)}
\affiliation{Key Laboratory for Research in Galaxies and Cosmology, Shanghai Astronomical Observatory, Chinese Academy of Sciences, 80 Nandan Road, Shanghai 200030, China}

\author[0000-0003-3713-2640]{Songmei Qin (秦松梅)}
\affiliation{Key Laboratory for Research in Galaxies and Cosmology, Shanghai Astronomical Observatory, Chinese Academy of Sciences, 80 Nandan Road, Shanghai 200030, China}
\affiliation{School of Astronomy and Space Science, University of Chinese Academy of Sciences, No. 19A, Yuquan Road, Beijing 100049, China}

\author[0000-0003-1864-8721]{Tong Tang (唐通)}
\affiliation{Key Laboratory for Research in Galaxies and Cosmology, Shanghai Astronomical Observatory, Chinese Academy of Sciences, 80 Nandan Road, Shanghai 200030, China}
\affiliation{School of Astronomy and Space Science, University of Chinese Academy of Sciences, No. 19A, Yuquan Road, Beijing 100049, China}

\author[0000-0002-4907-9720]{Li Chen (陈力)}
\affiliation{Key Laboratory for Research in Galaxies and Cosmology, Shanghai Astronomical Observatory, Chinese Academy of Sciences, 80 Nandan Road, Shanghai 200030, China}
\affiliation{School of Astronomy and Space Science, University of Chinese Academy of Sciences, No. 19A, Yuquan Road, Beijing 100049, China}

\author{Jinliang Hou (侯金良)}
\affiliation{Key Laboratory for Research in Galaxies and Cosmology, Shanghai Astronomical Observatory, Chinese Academy of Sciences, 80 Nandan Road, Shanghai 200030, China}
\affiliation{School of Astronomy and Space Science, University of Chinese Academy of Sciences, No. 19A, Yuquan Road, Beijing 100049, China}

\begin{abstract}

We investigated the stellar mass function and the binary fraction of 114 nearby open clusters (OCs) using the high-precision photometric data from {\it Gaia} Data Release 3 ({\it Gaia} DR3). We estimated the mass of member stars by using a ridge line (RL) that is better in line with the observed color-magnitude diagram (CMD), thus obtaining more accurate stellar mass and binary mass ratio ($q$) at the low-mass region. By analyzing the present-day mass function (PDMF) of star clusters, we found that 108 OCs follow a two-stage power-law distribution, whereas 6 OCs present a single power-law PDMF. Subsequently, we fitted the high(low)-mass index of PDMF ($dN/dm \propto m^{-\alpha}$), denoted as $\alpha_{\rm h}$($\alpha_{\rm l}$), and segmentation point $m_{\rm c}$. For our cluster sample, the median values of $\alpha_{\rm h}$ and $\alpha_{\rm l}$ are 2.65 and 0.95, respectively, which are approximately consistent with the initial mass function (IMF) results provided by \cite{2001MNRAS.322..231K}. We utilized the cumulative radial number distribution of stars with different masses to quantify the degree of mass segregation. We found a significant positive correlation between the state of dynamical evolution and mass segregation in OCs. We also estimated the fraction of binary stars with $q \geq 0.5$, ranging from 6\% to 34\% with a median of 17\%. Finally, we provided a catalog of 114 nearby cluster properties, including the total mass, the binary fraction, the PDMF, and the dynamical state.

\end{abstract}


\section{Introduction} \label{sec:intro}

Initial mass function (IMF) describes the mass distribution of stars that form together at the birth time. It is a crucial parameter in astrophysics \citep{2022A&A...664A.145D}, influencing most observable properties of star clusters and galaxies. The IMF serves as a critical input parameter for theoretical models or for deriving other physical parameters based on observations \citep{2005A&A...430..491R,2022ApJ...932..103G}. However, whether the IMF is universal or varies with the interstellar environment is still an open question \citep{2023arXiv230715831E}. Studying the IMF significantly contributes to understanding the star formation process \citep{2009ARA&A..47..371T,2015MNRAS.448.1847H}, as well as galaxy formation and evolution \citep{2016ApJ...824...82C}.

An open cluster (OC) is a group of coeval stars formed in a giant molecular cloud in less than $3\,\rm Myrs$ \citep{2003ARA&A..41...57L}. Being single stellar populations, young OCs maintain their initial stellar formation state within a dozen million years, which are widely used to study the stellar mass function \citep{2010ARA&A..48..339B}. Although the massive member stars would evolve away from the main sequence on the Hertzsprung-Russell diagram, most medium-mass and low-mass member stars are still ideal tracers for investigating the cluster's properties, such as the mass function and mass segregation. The coeval nature of an OC allows us to determine its members by their clustering in the position space, the proper motion distribution, and the color-magnitude diagram (CMD) \citep{2005A&A...440..403K}. Subsequently, one can accurately determine the age and mass of member stars by adopting the isochrone fitting on the CMD. In addition, all members in an OC share a similar distance and extinction, whose observability is only affected by intrinsic brightness. This results in a higher completeness of star counting within OCs than field stars.

It is worth noting that binaries are widely present in stellar systems with different masses. Therefore, the estimation of total mass and mass function considerably depends on the binary fraction \citep{2018arXiv180610605K,2019ApJ...874..127B,2020ApJ...896..152R}. When an unresolved binary is regarded as a high-mass single star, it will lead to deviations in constituting the mass function \citep{1993MNRAS.262..545K}. Besides, the difference in the initial binary fraction in clusters could also be misinterpreted as the variation of IMF \citep{2014MNRAS.442L...5M}. Since unresolved binaries of OCs located in brighter and redder areas of the CMD than single member stars \citep{1974A&A....32..177M, 1998MNRAS.300..977H},  it is easier to identify those binary stars on the CMD, which makes the estimation of binary fraction and mass function in OCs more reliable compared to those derived from field stars.

Several papers have reported on the properties of binary stars in OCs, including the binary fraction and the distribution of binary mass ratio ($q$). \citet{2020ApJ...903...93N} employed synthetic CMD to determine the binary fractions and fundamental parameters for 12 OCs. In their study, the binary fraction varies from 29\% to 55\%, and the distribution of $q$ in these OCs is flat, ranging from 0 to 1. \citet{2021AJ....162..264J} estimated the fraction of binaries with $q>0.6$ for 23 OCs. The binary fraction of stars with mass from $0.4\,\rm M_\odot$ to $3.6\,\rm M_\odot$ ranges from 12\% to 38\%, and reaches the peak between 12\% and 20\%, but it decreases with stellar mass. Furthermore, they suggested that the distribution of $q$ is unlikely to be flat. \citet{2022ApJ...930...44L} measured the properties of 10 OCs by assuming that $q$ follows a power-law distribution with a mixture model called MiMO. Their results indicate that the binary fraction ranges from 30\% to 50\% for binary stars with $q>0.2$. \citet{2023arXiv230111061D} estimated the fraction of binary stars with $q>0.6$ for 202 OCs within $1.5\,\rm kpc$ from the sun. The binary fraction of these OCs varies from 5\% to 67\%, with a median value of 15\%. Additionally, they found that the binary fraction increases with the mass of the primary star and decreases with the cluster abundance.

{\it Gaia} Data Release 3 ({\it Gaia} DR3) published unprecedented astrometric and photometric data for approximately 1.5 billion sources \citep{2023A&A...674A...1G}. The astrometric parameters (e.g., position, proper motion, and parallax) of {\it Gaia} DR3 are more precise than {\it Gaia} Data Release 2 ({\it Gaia} DR2). The median uncertainty in parallax is $0.02-0.03\,\rm mas$ at \ensuremath{G} = $9-14\,\rm mag$ and $0.5\,\rm mas$ at \ensuremath{G} = $20\,\rm mag$. Similarly, the median uncertainty of proper motion is $0.02-0.03\,\rm mas\,yr^{-1}$ for \ensuremath{G} = $9-14\,\rm mag$ and $0.5\,\rm mas\,yr^{-1}$ for \ensuremath{G} = $20\,\rm mag$ \citep{2021A&A...649A...2L}. Furthermore, it includes high-precision photometric parameters, e.g. magnitudes of \ensuremath{G}, \ensuremath{G_\mathrm{BP}}, and \ensuremath{G_\mathrm{RP}}, with an uncertainty of $0.2\,\rm mmag$ at \ensuremath{G} = $10-14\,\rm mag$ and $0.8\,\rm mmag$ at $\ensuremath{G} \approx 17\,\rm mag$ \citep{2021A&A...649A...3R}. With the help of high-quality {\it Gaia} data, the investigation of OCs has entered a new era. 

Besides the {\it Gaia} data, the growing development of membership determination algorithms also facilitated the discovery of more OCs. \citet{2018A&A...618A..93C} applied the UPMASK (Unsupervised Photometric Membership Assignment in Stellar Clusters, \citet{2014A&A...561A..57K}) algorithm with {\it Gaia} DR2 to identify the members of 1229 clusters, including 60 newly discovered OCs; \citet{2020A&A...635A..45C} used DBSCAN (Density-Based Spatial Clustering of Applications with Noise, \citet{1996kddm.conf..226E}) algorithm to search for overdensities in {\it Gaia} DR2 and successfully found 582 new OCs; in the study of \citet{2019ApJS..245...32L}, 2243 OCs were detected with {\it Gaia} DR2 by utilizing the FoF (Friends of Friends, \citet{1982ApJ...257..423H}) algorithm, which reveal 76 new OC candidates; combining the DBSCAN and pyUPMASK clustering algorithms in five-dimensional phase space ($d_{l^*}$, $d_b$, $v_{\alpha^*}$, $v_{\delta}$, $\varpi$), \cite{2022ApJS..260....8H} carried out a blind search using {\it Gaia} data, and eventually found 541 unreported star clusters. In addition, more OCs are confirmed with extended halo \citep{2019A&A...621L...3M, 2019A&A...621L...2R, 2020ApJ...889...99Z, 2019A&A...624A..34Z, 2022AJ....164...54Z, 2022RAA....22e5022B}, improving the completeness of cluster members. It provides a valuable opportunity for a comprehensive understanding of star clusters and critical data for better studying the mass function and binary fraction in OCs.

In the {\it Gaia} era, the investigation of stellar mass function in OCs has been advanced using updated membership identification results. \citet{2022MNRAS.517.3525B} estimated the member star masses for 46 OCs with extended structure using the {\it Gaia} Early Data Release 3 ({\it Gaia} EDR3) and fitted the mass functions of their members ($>0.5\,\rm M_\odot$) in the intra-tidal, extra-tidal, leading tail, and trailing tail for each OC. They found that most older clusters are usually mass-segregated and present a flatter mass function. In {\it Gaia} EDR3, \citet{2022MNRAS.516.5637E} analyzed the mass functions of 15 nearby OCs. They fitted the present-day mass function (PDMF) with a single power law and found the slopes are $-3<\alpha<-0.6$. In particular, they evidenced a significant correlation between the PDMF slope and the ratio of age to half-mass relaxation time. \citet{Cordoni_2023} used {\it Gaia} DR3 to determine the PDMF by converting magnitudes into mass for 78 OCs. They fitted the mass function for the low- and high-mass end, respectively, with the mass segmentation point at $1\,\rm M_\odot$. In the study of \citet{2023MNRAS.525.2315A}, stellar masses were derived for 773 OCs using synthetic clusters through the Monte Carlo method, and then a two-part segmented linear function was employed to fit the PDMF.

\defcitealias{2023ApJS..265...12Q}{Qin23}

Using the {\it Gaia} DR3, \citet[hereafter \citetalias{2023ApJS..265...12Q}]{2023ApJS..265...12Q} developed a slicing approach to search for OCs in the solar vicinity within $500\,\rm pc$. Their published catalog of 324 OCs includes an updated membership of previously reported OCs and 101 newly discovered OCs. This increases the number of discovered OCs in the solar vicinity by $\approx45\%$. For the neighboring OCs provided by \citetalias{2023ApJS..265...12Q}, considering that their membership probability is more reliable than previous studies, and the completeness of low-mass stars is high, we decided to perform a detailed analysis (mass function and binary fraction) for these OCs. 

Since cluster members share similar age and chemical abundance, the mass of members can be easily determined from theoretical isochrones \citep{1992ApJS...81..163B}. This paper mainly aims to estimate the stellar mass and the binary fraction of 114 OCs in the solar vicinity within $500\,\rm pc$. In Sect.~\ref{sec:sam}, we introduce the selection criteria for the OC sample. In Sect.~\ref{sec:met}, we describe a revised approach for isochrone fitting, stellar mass determination, and binary fraction estimation using ridge line (RL). In Sect.~\ref{sec:res}, we present the results and discuss the PDMF, mass segregation, and binary faction of 114 OCs. Finally, we summarize this work in Sect.~\ref{sec:sum}.

\section{Sample} \label{sec:sam}

\citetalias{2023ApJS..265...12Q} provides a catalog of 324 OCs in the solar neighborhood within $500\,\rm pc$, including 101 new OCs and 223 previously reported OCs. In the {\it Gaia} era, it is confirmed that star clusters generally have an extended structure, while their physical scales are believed to be approximately $10-100\,\rm pc$ \citep{2022AJ....164...54Z}. To obtain more complete member stars in an OC,  \citetalias{2023ApJS..265...12Q} expanded the search area (with a radius of $50\,\rm pc$) for member star identification. The OCs in \citetalias {2023ApJS..265...12Q} indeed have a larger physical size than the previous catalog \citep{2020A&A...640A...1C}, which indicates that the stellar mass function of OCs based on the member star catalog of \citetalias {2023ApJS..265...12Q} would be more reliable than previous studies.

In \citetalias{2023ApJS..265...12Q}, an approach called pyUPMASK \citep{2021A&A...650A.109P} is adopted to perform the membership identification. The pyUPMASK is an unsupervised clustering method for stellar clusters, developed in Python based on the original UPMASK\citep{2014A&A...561A..57K}, to determine membership probabilities. This makes it more versatile, improves its performance, and further provides a variety of core algorithms for selection. It is a robust method successfully employed in numerous studies for identifying cluster membership \citep{2018A&A...618A..93C,2019A&A...624A.126C,2020A&A...640A...1C}. In particular, \cite{2022RAA....22e5022B} estimated the completeness and contamination of pyUPMASK in identifying member stars by adopting ten simulated clusters. Their result shows that the average identification completeness and contamination of ten simulated clusters are about 97\%  and 4\%, respectively.

\begin{figure}[ht!]
\centering
%\plotone{CC.pdf}
\includegraphics[width=0.6\textwidth, angle=0]{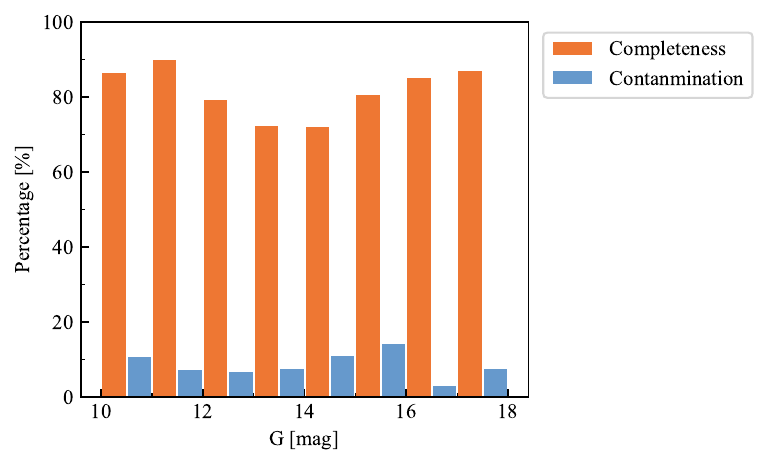}
\caption{Completeness (orange) and contamination (blue) rate as a function of \ensuremath{G} magnitude. The completeness and contamination rate median values are approximately 83\% and 8\%, respectively.
\label{fig:CC}}
\end{figure}

However, since the ratio of member stars ($N_m$) to field stars ($N_f$) is crucial for identifying member stars with pyUPMASK, it is essential to reassess the completeness and contamination in \citetalias{2023ApJS..265...12Q} cluster sample. We first counted the number of member stars and field stars in \citetalias{2023ApJS..265...12Q} cluster sample and found their median values were 235 and 80, respectively. Then, we selected ten OCs whose ($N_m$, $N_f$) are close to (235, 80) and adopted their observational parameters (e.g., spatial position, proper motion, parallax, and \ensuremath{G} magnitude) as the input parameters for the simulation. For each simulated cluster field, we employed pyUPMASK to assign the membership probability ($p$), and like \citetalias{2023ApJS..265...12Q}, we regard stars with $p>0.5$ as member stars, while others are field stars. Lastly, the completeness and contamination of membership identification can be estimated by comparing the input simulated cluster field and the pyUPAMSK result. We combined the results of ten simulated cluster fields to further investigate the completeness and contamination as a function of the \ensuremath{G} magnitude. We divided their \ensuremath{G} magnitude into eight bins. The completeness and contamination rate for each bin is shown in Figure~\ref{fig:CC}. The median values of completeness and contamination rate in our simulated cluster fields are approximately 83\%  and 8\%, respectively, which illustrate that the cluster members in \citetalias{2023ApJS..265...12Q} have the properties of relatively high completeness and low contamination rate.

The number of OC members varies greatly ($10^2-10^4$), and most low-mass OCs have undergone or are undergoing dissolution, which leads to a significant difference between the PDMF and the IMF. In addition, for OCs with a low number of stars, the obtained PDMF is unreliable due to low statistics. After excluding low-mass OCs whose members are less than 100, we first selected 180 out of 324 OCs as the initial sample. Considering the broadening effect of the main sequence (MS) caused by the differential reddening may bias the estimation of total mass and binary fraction, we further excluded clusters suffering from broadening of the MS by visual inspection. Finally, 114 OCs (17 of which were reported by \citetalias{2023ApJS..265...12Q} for the first time) were retained as high-quality clusters with sufficient member stars and clear MS to perform the stellar mass estimation.

\begin{figure}[ht!]
\centering
\includegraphics[width=1\textwidth, angle=0]{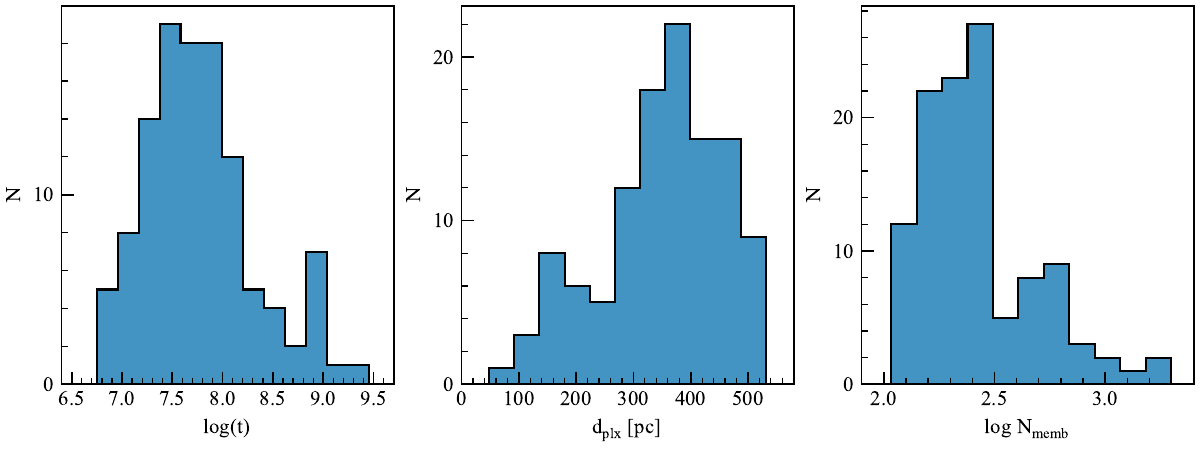}
\caption{Histogram of cluster age (left), distance (middle), and number of member stars (right) of 114 OCs in our sample.
\label{fig:dist}}
\end{figure}

In Figure~\ref{fig:dist}, we presented the distribution of age, distance, and the number of members of 114 high-quality OCs. Although our cluster sample covers a wide age range ($\log(t)=6.750-9.450$), most of them are relatively young ($\log(t)< 8.500$) with a peak at $\log(t)=7.500$. The distribution of distances, inferred from the averaged parallax of members, indicates that most OCs are within $500\,\rm pc$, peaking at $400\,\rm pc$. The number of stars in most OCs is less than 300. It is noteworthy that there are four massive clusters with a large number of members ($>1000$) in our sample, namely Trumpler 10, Melotte 22, NGC 2516, and NGC 3532, which may become an important sample for further studying the formation and evolution of star clusters. 

%%%%%%%%%%%%%%%%%%%%%%%%%%%%%%%%%%%%%%%%%%%%%%%%%%%%%%%%%%%%%%%%%%%%%%%%%%%%%%%%%%%%%%%%
\section{Method} \label{sec:met}

The standard method for determining the mass of a member star is to correspond its intrinsic color and absolute magnitude to the theoretical isochrone and then infer its mass from the theoretical mass of the isochrone \citep{2021MNRAS.505.1607B,2021AJ....161....8Y}. Since unresolved binaries are usually redder and brighter than the MS of single stars on the CMD \citep{2021AJ....162..264J}, the primary mass and $q$ of binary stars can also be derived by constructing the MS of binary stars. This method is effective for clusters with a clear MS and easily fitting isochrone on the CMD. Hence, reliable isochrone fitting is critical for determining stellar mass in a cluster.

\begin{figure}[ht!]
\centering
%\plotone{deviation.pdf}
\includegraphics[width=0.8\textwidth, angle=0]{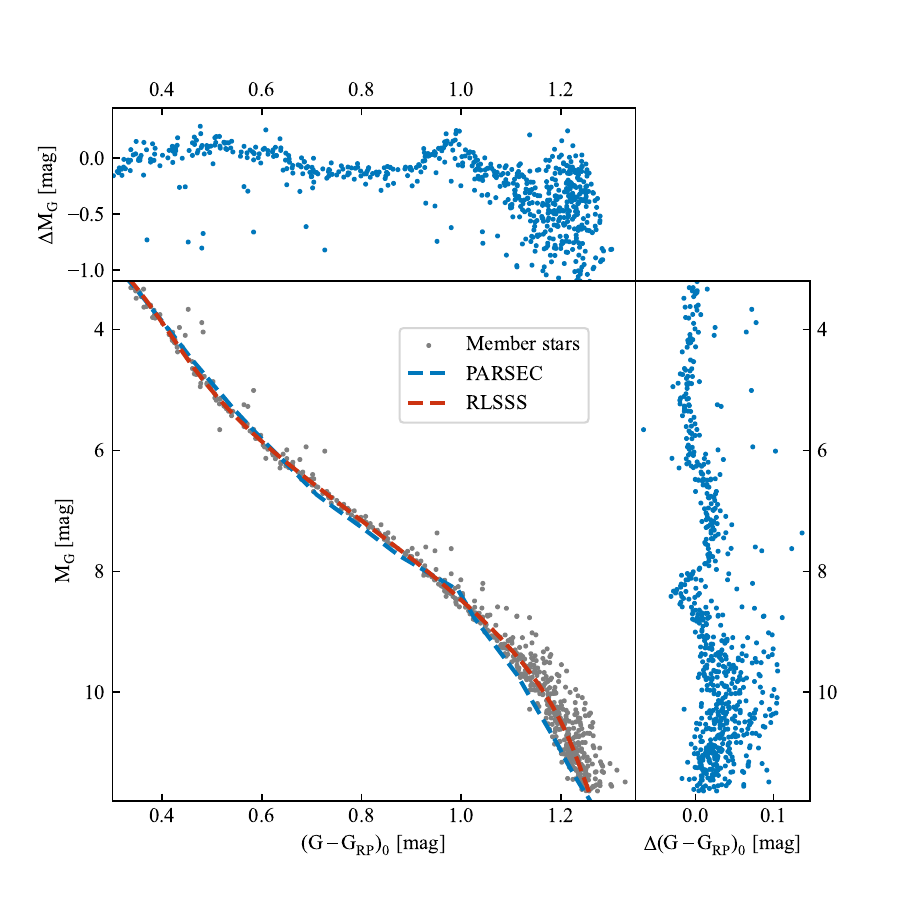}
\caption{CMD of member stars in OCSN 219. Grey dots represent member stars, while red and blue dashed lines correspond to the best results of the RLSSS and isochrone, respectively. The right and upper panels show the difference between observed data and the optimal isochrone in magnitude and color space, which displays a significant offset and dispersion at the low-mass end.
\label{fig:devi}}
\end{figure}

In the isochrone fitting process, the distribution of MS turn-off stars and giant stars depends on cluster age. Therefore, bright stars commonly play a significant role in isochrone fitting. However, due to the large number of low-mass stars in the cluster, if the Least-squares fitting is performed on all member stars with equal weights, it will result in excessive fitting weights at the low-mass end of the cluster and bias in the age estimation. Therefore, different fitting weights need to be carefully assigned based on the magnitude distribution of the cluster. In addition, despite the distribution of bright stars and isochrone being in good agreement, there are significant differences at the low-mass ends of the observed data for some star clusters \citep{2019A&A...622A.110F,2020ApJ...901...49L,2023AJ....165..108B}. This is due to uncertainties in modeling the atmospheres of very low-mass stars, as they are cold enough to form molecules and even dust in some cases. This deviation would induce a significant error and uncertainty for stellar mass determination, especially for binaries. For example, Figure~\ref{fig:devi} presents a cluster's distribution of member stars on the CMD, which clearly shows the significant difference between the isochrone and the observed data at the low-mass end.

Single stars with the same mass will exhibit similar color and brightness, resulting in a high-density observed sequence (RL) on the CMD. By better reflecting the distribution characteristics of the MS of single stars, the RL enhances the distinction between single and binary stars, reduces the fitting deviation of the MS of single stars due to extended MS effects, and ultimately improves the accuracy of the isochrone fitting. On the other hand, if all observed stars have the same fitting weight, the low-mass end with more stars will have a greater fitting weight, resulting in the isochrone fitting being entirely dominated by the low-mass end. This shortage can be partly overcome by weighting the luminosity of stars. However, due to the varying proportions of high- and low-mass stars in different star clusters, inappropriate luminosity weighting may also lead to a significant deviation in the isochrone fitting result. Alternatively, when utilizing the RL with the same magnitude interval instead of the observed data in isochrone fitting, the fitting weights for both low-mass and high-mass ends are the same, entirely overcoming the shortcoming that the fitting weight for the low-mass end is affected by the number of stars, resulting in more reasonable fitting results.

The main steps of our method are listed as follows:
\begin{enumerate}
    \item[(1)] Using Robust Gaussian Process Regression based on Iterative Trimming (ITGP) \footnote{\url{https://github.com/syrte/robustgp}} to determine the RL of single star sequence (RLSSS).
    
    \item[(2)] Fitting the isochrone with RLSSS to obtain the parameters, e.g., age: $\log(t)$; metallicity: $Z$; distance modulus: $DM$; reddening: $E(B-V)$.
    
    \item[(3)] Estimating the stellar masses and the $q$ of binary stars from the CMD-based RL matching method.
\end{enumerate}

\subsection{Ridge Line} \label{subsec:rid}

ITGP is a new robust regression method based on standard Gaussian Process (GP) and iterative trimming, which is easy to implement and computationally simple \citep{2021A&C....3600483L}. ITGP first runs the standard GP in all members to get a rough RLSSS in terms of color as a function of magnitude, then removes a fraction of outliers (binary stars) farthest from the RLSSS in color and re-runs the GP to update the RL iteratively. This approach introduces a novel shrinking procedure that gradually increases the trimming fraction to prevent premature convergence and a one-step re-weighting to improve the statistical efficiency. It is noted that \citet{2021A&C....3600483L} provides a set of recommended parameters and proves that for general problems, it can determine the optimal RLSSS well after several iterations.

\begin{figure}[ht!]
\centering
\includegraphics[width=0.8\textwidth, angle=0]{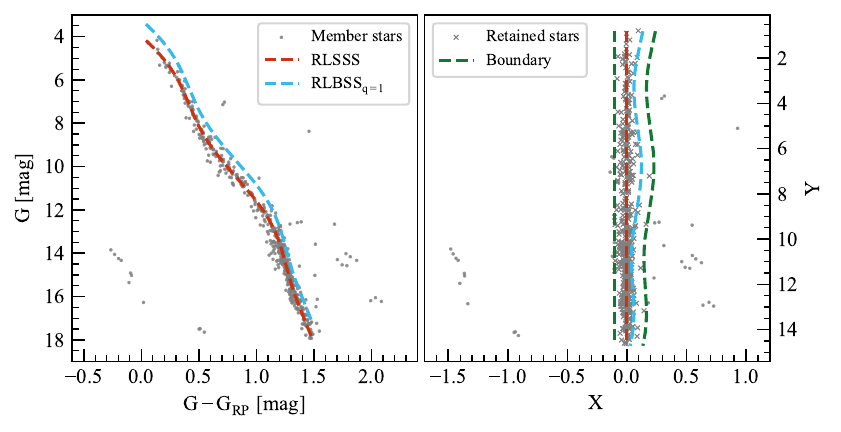}
\caption{Left panel: CMD of OCSN 220 (gray dots). The red and blue dashed lines represent the RLSSS and $\textup{RLBSS}_{q=1}$ respectively, which effectively trace the distribution of member stars. Right panel: distribution of member stars on the $\rm X-Y$ plane after converting coordinates along the RLSSS. Besides the RLSSS and $\textup{RLBSS}_{q=1}$ dashed lines same to the left panel, the newly added two green dashed lines represent two boundaries ($\rm RLSSS-3\sigma_m$ and $\textup{RLBSS}_{q=1}+3\sigma_m$) respectively. Within this range, the member stars used for stellar mass estimation are represented by a cross symbol. It is worth noting that the position of white dwarfs in the lower left corner of the two panels has little changed. This is because their shortest connection lines to the RLSSS are almost parallel to the $\ensuremath{G}-\ensuremath{G_{\rm RP}}$ axis, which means the perpendicular distance between the white dwarfs and the RLSSS is mainly contributed by the $\ensuremath{G}-\ensuremath{G_{\rm RP}}$ colors, so the relative positions of the white dwarfs do not change significantly in the two coordinates.
\label{fig:X-Y}}
\end{figure}

In Figure~\ref{fig:X-Y}, according to the distribution of members on the CMD, we first adopted the parameters recommended by ITGP to derive an optimal RLSSS and added 0.75 mag to acquire an RL of binary stars sequence (RLBSS) with $q=1$ ($\textup{RLBSS}_{q=1}$). Then, to obtain a purer sample for mass estimation, we exclude possible field stars and white dwarfs on the CMD according to the separation between each star and the RL. To achieve this, we converted all the observation data and the $\textup{RLBSS}_{q=1}$ to the coordinates (X, Y). In this coordinate system, the X-axis represents the distance perpendicular to RLSSS, with the point bluer than RLSSS being negative and redder than RLSSS being positive; the Y-axis is the distance traveled along RLSSS from zero point, with the brightest point of RLSSS being zero point. Along with the Y-axis, we divided the MS stars into bins ($\rm bin\_width = 1$) and calculated the median value of the standard deviations of the X in each bin ($\sigma_m$). The selection criterion is between the RLSSS minus 3$\sigma_m$ and the $\textup{RLBSS}_{q=1}$ plus 3$\sigma_m$. After excluding stars outside this boundary region, a purer sample including high-probability member stars was obtained to estimate the stellar mass function of the cluster. 

\subsection{Isochrone Fitting} \label{subsec:iso}

\citetalias{2023ApJS..265...12Q} provided the age, reddening, and distance modulus parameters for more than 300 OCs by visual fitting. Considering that the estimation of the stellar mass and the binary fraction is significantly dependent on the result of the isochrone fitting, we used the fitting result of \citetalias{2023ApJS..265...12Q} as the initial input parameters $\log(t)_{\rm ini}$, $DM_{\rm ini}$, $E(B-V)_{\rm ini}$ to perform the re-fitting procedure for the 114 OCs of our sample.

We downloaded the Padova isochrones \footnote{\url{http://stev.oapd.inaf.it/cgi-bin/cmd}} \citep{2012MNRAS.427..127B} with {\it Gaia} photometric system \citep{2021A&A...649A...3R}. For each star cluster, the parameter ranges and intervals (min, max, step) of the isochrones we used are $\log(t) \sim$ ($\log(t)_{\rm ini}-1$, $\log(t)_{\rm ini}+1$, 0.025) and $Z \sim$ (0.002, 0.04, 0.002), respectively. Besides, the other two fitting parameters $DM \sim$ ($DM_{\rm ini}-1$, $DM_{\rm ini} + 1$, 0.1) and $E(B-V) \sim$ ($E(B-V)_{\rm ini}-0.5$, $E(B-V)_{\rm ini}+0.5$, 0.01) were adopted to perform the isochrone fitting process. It is noticed that we used the $A_{G} = 2.74 \times E(B-V)$ and $E(G-RP)=0.705 \times E(B-V)$ \citep{2018MNRAS.479L.102C,2019A&A...624A..34Z} for the transformation of extinction, where $A_G$ is the average extinction of \ensuremath{G} band.

After determining the RLSSS of a cluster (See Sect.~\ref{subsec:rid}), we traversed through four fitting parameters to fit isochrone with RLSSS. At first, we interpolated the Padova isochrones based on the range and number of data points on RLSSS. Then, for each data point on RLSSS, we used the {\it KDTree} \footnote{\url{https://docs.scipy.org/doc/scipy/reference/generated/scipy.spatial.KDTree.html}} \citep{bentley1975multidimensional} to calculate the minimum distance to each isochrone corresponding to a set of fitting parameter ($\log(t)$, $Z$, $DM$, $E(B-V)$). Third, we summed the minimum distances of all RLSSS data points as the characteristic distance ($d_i$) of the corresponding parameter combination. At last, after traversing all the fitting parameter combinations, the combination with the smallest $d_i$ is considered as the optimal fitting parameter. After these processes, we finally determined the best-fitting isochrone of each cluster. 

Figure~\ref{fig:devi} presents the CMD of a cluster (OCSN 219) as well as the optimal isochrone fitting result. Although the bright end ($M_G<7$) of the optimal isochrone is consistent with the observed CMD, there is a significant offset at the low-mass end ($M_G>7$). In contrast, due to the better alignment of the RLSSS with the distribution of observational points, more reasonable measurements of member star mass and $q$ can be obtained.

\subsection{Mass Estimation} \label{subsec:mass}

In this work, we focused on determining the masses of pre-main sequence (PMS) and MS stars, so we exclusively retained the isochrone points with label = 0, 1, corresponding to the stellar evolutionary stages of PMS and MS respectively \citep{2017ApJ...835...77M}. As shown in Figure~\ref{fig:devi}, since even the optimal isochrone can not match well with the observation data of member stars, we assigned the data points of the optimal isochrone to the RLSSS based on the principle of the minimum distance. As a result, we generated the RLSSS, which includes information such as mass, luminosity, and color, which can be utilized to estimate stellar mass through the CMD.

In binary stars with different $q$, $m_2$ can be calculated from $m_2=m_1*q$, and then the corresponding magnitude of the secondary star can be derived by interpolating the RLSSS. Therefore, the RLBSS with different $q$ can be synthesized. In particular, it is difficult to distinguish the low-$q$ unresolved binaries from single stars because of the effects of observation uncertainties and differential reddening. Hence, we regarded binary stars with $q < 0.5$ as single stars. Eventually, we synthesized a set of RLBSS with $q$ from 0.5 to 1 and an interval of 0.1 for the mass estimation of binary stars. We considered the RLBSS with $q=0.5$ ($\textup{RLBSS}_{q=0.5}$) as the dividing line between single stars and binary stars.

\begin{figure}[ht!]
\centering
%\plotone{binaries.pdf}
\includegraphics[width=0.55\textwidth, angle=0]{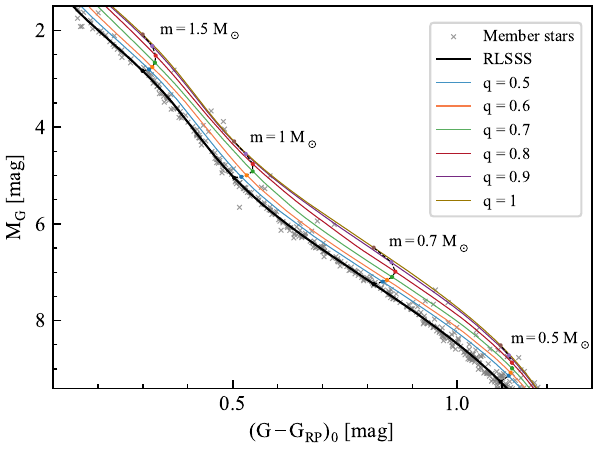}
\caption{RLSSS and RLBSS ($q \geq 0.5$) of a cluster (OCSN 219) on the CMD. Grey crosses represent member stars. Colorful solid lines show the RLBSS with $q =$ 0.5, 0.6, 0.7, 0.8, 0.9, 1. In addition, we used the black dashed lines to mark the binaries distribution with a mass of primary star $m = 0.5, 0.7, 1, 1.5\,\rm M_\odot$ for different $q$. These synthetic ridge lines effectively divided the major distribution area of member stars into a consistent grid, which makes the difference between the member star and the nearest point on the synthetic RL tiny. This also indicates that estimating the stellar mass by matching it with the synthetic RL is reliable.
\label{fig:binar}}
\end{figure}

Figure~\ref{fig:binar} shows the RLSSS and RLBSS ($q \geq 0.5$) of a cluster (OCSN 219) on the CMD. The major distribution area of members is divided into different grids by these synthetic ridge lines effectively. This allows us to easily match the member stars with the nearest points in the synthetic RL with minimal distance. So the mass of stars below the $\textup{RLBSS}_{q=0.5}$ is determined by matching with the RLSSS, and the mass and $q$ of stars above the $\textup{RLBSS}_{q=0.5}$ are calculated by matching with the nearest RLBSS.

\subsection{Verifying Stellar Mass} \label{subsec:comp}

To verify the mass estimated by the synthesized RL matching compared to the isochrone matching, we applied both matching methods to derive the stellar masses for the same cluster (OCSN 234), respectively. Since the difference between the isochrone and the RL mainly occurs at the low-mass end (as shown in Figure~\ref{fig:devi}), only the members with mass less than $\rm 0.5\,\rm M_\odot$ are taken into account. Furthermore, we cross-matched the members with the CTL (Candidate Target List) catalog to procure the mass provided by the TESS (Transiting Exoplanet Survey Satellite) mission \citep{2019AJ....158..138S} for comparison. TESS is an all-sky photometric survey conducted over two years with the goal of discovering exoplanets that transit around bright host stars. The CTL catalog is a comprehensive collection source for TESS 2-min cadence observations. The stellar mass in the CTL catalog is calculated from the empirical mass-luminosity relation ($\rm 0.075 - 0.7\,\rm M_\odot$) \citep{2019A&A...621L...3M}, which is created based on the orbital parameters of 62 binary stars. This is a different approach for estimating stellar mass; accordingly, it can be used to effectively assess the two CMD-based mass matching results as an external comparison.

\begin{figure}[ht!]
\centering
\includegraphics[width=0.8\textwidth,align=c]{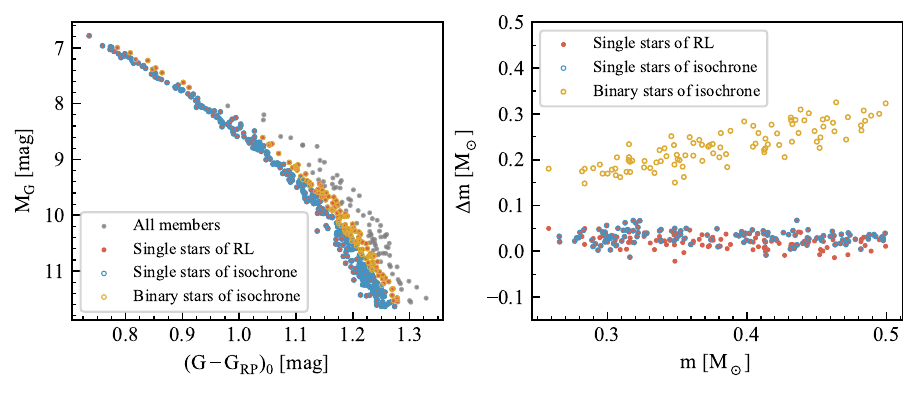}
\caption{Left panel: CMD distribution of single stars of OCSN 234. The red dots and blue circles represent single stars identified by the RL and isochrone matching methods, respectively. In comparison, about 28\% of stars (130 stars) were identified as single stars by the RL method but as binary stars by the isochrone method, which were present as yellow circles. Right panel: comparison of stellar masses based on the two matching methods, with the TESS mass as a reference. The stellar masses of yellow circles are the total mass ($m_1+m_2$) of the binary stars identified by the isochrone method. For these stars, because their $q$ changes from 0.5 to 1, this results in a linear distribution of total mass along the mass axis. In contrast, single stars (red dots) identified by the RL method present a more reliable mass result due to the small dispersion result.}
\label{fig:delta_m}
\end{figure}

Figure~\ref{fig:delta_m} compares single stars results identified by the RL and isochrone matching methods. Due to the difference in the distribution of the single-star MS used by the two methods, the RL method identifies more stars as single stars. On the other hand, some of the single star members from the RL method are recognized as binary stars (yellow circles) by the isochrone method. This is because they are far from the single-star MS of isochrone (see Figure~\ref{fig:devi}). Furthermore, we compared the differences in masses for stars between the two CMD-based matching methods and TESS. For single stars recognized by both methods, their mass is consistent with the mass of TESS, with a systematic offset of $0.031\,\rm M_\odot$ and a dispersion of $0.014\,\rm M_\odot$. For stars identified as single stars by the RL method but binary stars by the isochrone method, their total mass ($m_1+m_2$) from the isochrone method exhibits a significant bias with the TESS mass. It can be seen that using the isochrone method will result in some single stars being mistakenly classified as binary stars, thereby overestimating their stellar mass. Therefore, the RL method presents a more reliable mass estimation result for single stars than the isochrone method.

We compared numbers of different $q$ of two CMD-based mass matching results in Table~\ref{tab:q}. Similarly, owing to the offset between the isochrone and the observed CMD (or synthesized RL), the actually observed magnitude of most low-mass stars is brighter than the magnitude of the binary stars with $q=1$ derived by the isochrone. This will result in most stars being identified as binary stars with $q=1$. As a comparison, the $q$ derived from the RL is reasonably distributed in the range of $0.5-1$ and consistent with the power law distribution $\chi_q(q) \propto q^\gamma$ \citep{2007A&A...474...77K,2013A&A...553A.124R,2013ARA&A..51..269D}. In summary, the RL mass matching method is more reasonable for estimating the mass and $q$ of binary stars.

\begin{table}[h]
  \centering
  \caption{The number of different $q$ determined by two methods.}
  \label{tab:q}
  \normalsize% fontsize
  \setlength{\tabcolsep}{8pt}% column separation
  \renewcommand{\arraystretch}{1.5}%row space 
  \begin{tabular}{ccccccc}
    \hline
    $q$ & 0.5 & 0.6 & 0.7 & 0.8 & 0.9 & 1.0 \\
    \hline
    $N_{\rm iso}$ & 0 & 0 & 0 & 0 & 7 & 58 \\
    $N_{\rm RL}$ & 6 & 5 & 6 & 9 & 9 & 30 \\
    \hline
  \end{tabular}
\end{table}

\section{Results} \label{sec:res}

\subsection{Mass Function} \label{subsec:PDMF}

The IMF in the Galactic stellar system is usually described as a single power law $dN/dm \propto m^{-\alpha}$ (Salpeter 1955: $\alpha=2.35$) \citep{1955ApJ...121..161S}, or a multiple-part power law variation with different mass ranges \citep{2001MNRAS.322..231K}, or a power law for $m \geq 1\,\rm M_{\odot}$ and a log-normal below $1\,\rm M_\odot$ \citep{2003PASP..115..763C}. Although the definitions of PDMF and IMF are different, for star clusters with coeval member stars, the evolution of stars will result in the loss of the mass function at the high-mass end without affecting the mass function at the low-mass end of the MS stars. However, because of the mass segregation, low-mass stars in the outer region are more likely to be stripped from the cluster by tidal shocking events \citep{2021PhDT........12T}, which results in a deviation from the IMF at the low-mass end. For OCs with shorter evolutionary histories, this effect has a very limited impact on the mass function. Therefore, we can consider the PDMF and IMF of the young OCs to be similar.

\begin{figure}[ht!]
\centering
%\plotone{mass_dist.pdf}
\includegraphics[width=1\textwidth, angle=0]{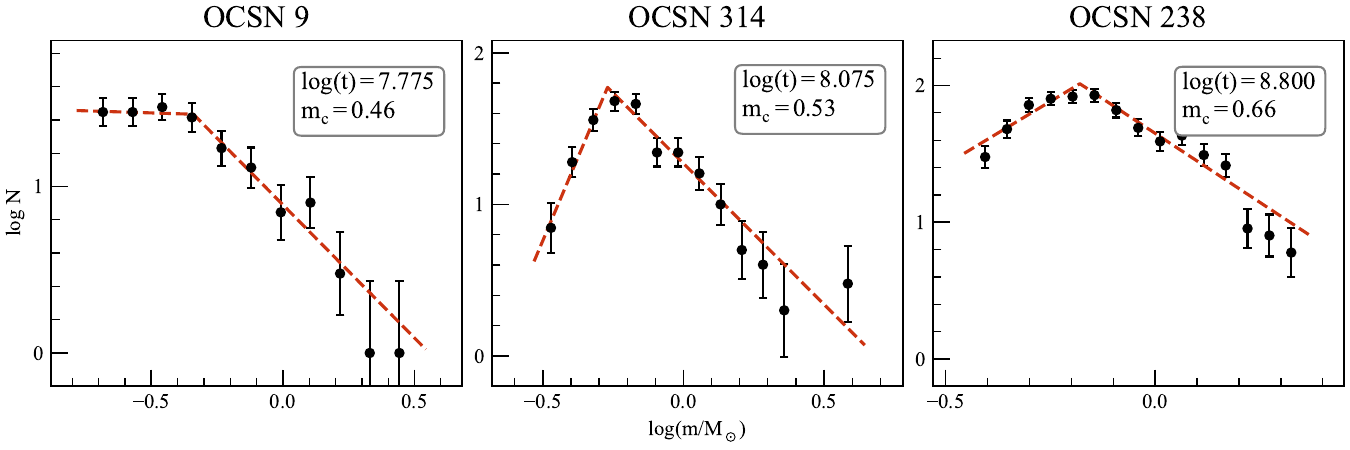}
\caption{Mass function of three clusters OCSN 9, OCSN 314, OCSN 238. The error bar is the Poisson noise of the number of stars in each mass bin.}
\label{fig:m_dist}
\end{figure}

As an example, we display the mass function of three clusters (OCSN 194, OCSN 314, OCSN 90) with different ages ($\log(t)$ = 7.225, 8.075, 8.300) in Figure~\ref{fig:m_dist}. When both the masses and the number of stars are represented in logarithmic scale, the mass function of most clusters was found to present a two-part segmented linear form. Given the star clusters get more dynamically evolved as they age, it follows that we observe varying degrees of difference between the PDMF and IMF at the low-mass end. To differentiate between mass functions within various mass ranges, we referred to the approach of \citet{2023MNRAS.525.2315A}\footnote{\url{https://github.com/ander-son-almeida/DashboardOCmass/blob/main/examples/MF_example.ipynb}} and employed a two-part segmented linear function to fit the mass function through the Least-squares algorithm, which is defined as:
\begin{equation}
    F(x) = \left\{
        \begin{array}{ll}
            -\alpha_{\rm h}x+b_1 & \text{if } x > m_{\rm c} \\
            -\alpha_{\rm l}x+b_2 & \text{if } x \geq m_{\rm c}
        \end{array}
        \label{equ:two}
    \right.
\end{equation}
where $\alpha$ and $b$ represent the slope of the function and the corresponding intercepts for the high- and low-mass end, respectively. $m_{\rm c}$ is the mass value of the segmentation point, which is also set as a free parameter in the fitting procedure.

\begin{figure}[ht!]
    \centering
    \includegraphics[width=0.4\textwidth,angle=0]{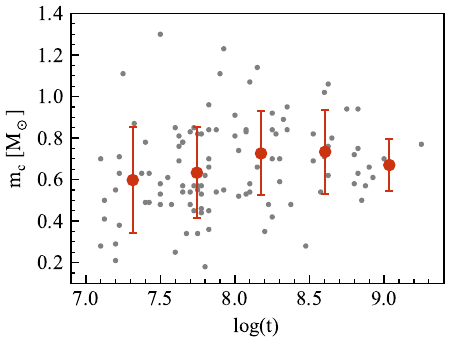}
    \caption{Segmentation point of a cluster as a function of cluster age. Clearly, the older the cluster, the greater the mass of the segmentation point. The interaction among member stars results in a significant number of low-mass stars being ejected, causing substantial modifications in the PDMF of the low-mass and becoming more pronounced as the cluster ages.}
    \label{fig:m_cut}
\end{figure}

\begin{figure}[ht!]
    \centering
    \includegraphics[width=0.6\textwidth,angle=0]{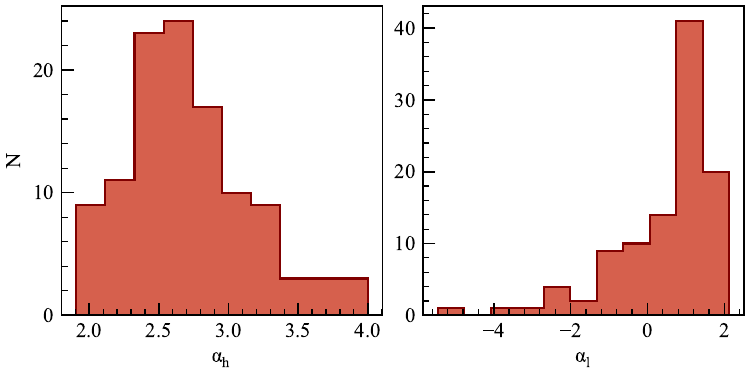}
    \caption{Histogram of power-law values of PDMF. The $\alpha_{\rm h}$ and $\alpha_{\rm l}$ represent the PDMF distribution with values greater than or less than the segmentation point, with median values of 2.65 and 0.95, respectively. To ensure the reliability of the PDMF result, we exclude the $\alpha_{\rm l}$ of 7 OCs since their number of mass bins is less than three at the low-mass end, which will result in insufficient statistical significance.}
    \label{fig:Gamma_dist}
\end{figure}

In our sample, we found that the PDMF of 108 OCs follows a two-segment power-law function. Four OCs only present a single power-law form at the high-mass end, and two OCs only present a single power-law form at the low-mass end. Figure~\ref{fig:m_dist} presents the fitting results of three example clusters. In addition, we exhibit the cluster's segmentation point ($m_{\rm c}$) as a function of cluster age in Figure~\ref{fig:m_cut}. It shows a weak relation that the older the cluster, the greater the mass of the segmentation point, which also indicates that PDMF may change with the cluster age. In Figure~\ref{fig:Gamma_dist}, we display two histograms of $\alpha_{\rm h}$ and $\alpha_{\rm l}$. The median values of $\alpha_{\rm h}$ and $\alpha_{\rm l}$ are 2.65 and 0.95, respectively, which is quite consistent with the multiple-part power-law IMF in \citet{2001MNRAS.322..231K}.

\begin{figure}[ht!]
%\plotone{G_plot}
\centering
\includegraphics[width=0.6\textwidth, angle=0]{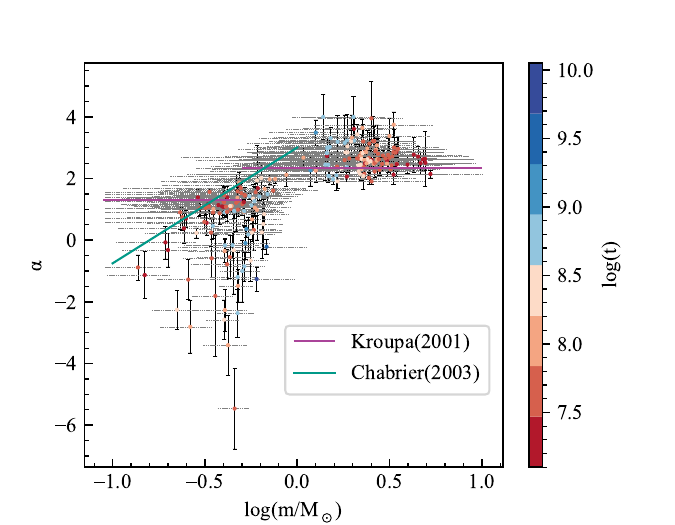}
\caption{The $alpha-plot$. The data show the derived $\alpha$ over different mass ranges determined by the segmentation point, while points are placed in the center of the mass range, and dashed lines indicate the mass range used to fit each $\alpha$. The violet solid line represents an IMF with multiple-part power-law form \citep{2001MNRAS.322..231K}, while the green line represents the IMF with a log-normal form \citep{2003PASP..115..763C}. The colors of each point represent the cluster age. In general, the PDMF of OCs is roughly consistent with the multiple-part power-law IMF in \citet{2001MNRAS.322..231K}. Furthermore, after excluding older star clusters that have undergone significant dynamic evolution and stellar evolution, the distribution of low-mass PDMF of OCs is more likely to be a lognormal pattern (red dots). 
\label{fig:G_plot}}
\end{figure}

Figure~\ref{fig:G_plot} presents the $\alpha$ distribution as a function of different mass ranges for 114 OCs, while colors represent the age of clusters. Although the segmentation points of each cluster are different, the $\alpha$ in the high-mass part can be roughly considered as constant, independent of the cluster age. We also note that the primary dispersion comes from older clusters, as more dynamic evolution effects influence their PDMF. After ignoring these older clusters with large dispersion, it is more apparent that the $\alpha$ of the high-mass end in PDMF for OCs is a constant, consistent with the value of Kroupa's IMF. In addition, for the low-mass part of the cluster, the difference between the PDMF and IMF of the cluster increases with age, resulting in a significant dispersion of the $\alpha$ distribution. However, after excluding older clusters, the $\alpha$ distribution of young clusters (red points) is consistent with Chabrier's IMF. It is worth noting that, for the low-mass end, the IMF variation may depend on the interstellar environments, such as the metallicity \citep{2023Natur.613..460L}. Considering the diverse formation environments of star clusters, this could result in a certain degree of dispersion in the $\alpha$ distribution of low-mass parts.


\subsection{Mass Segregation} \label{subsec:seg}

Many studies have found that a large number of OCs have the mass segregation phenomenon \citep{2018MNRAS.473..849D,2019A&A...629A.135D,2019MNRAS.490.2521H}, that compared with low-mass stars, massive stars tend to be distributed in the central region of the cluster. Mass segregation is generally believed to result from clusters' dynamic evolution. Member stars in the high number density environment of OCs have a greater probability of interacting with each other. When two stars encounter or pass by, energy equipartition causes low-mass stars to strive toward having the same kinetic energy as high-mass stars. Owing to their lower mass, they obtain higher velocity and exhibit more significant velocity dispersion, resulting in a more extensive spatial distribution. Meanwhile, the high-mass stars tend to exhibit smaller velocity and settle toward the cluster center \citep{2021PhDT........12T}. However, it is found that some clusters with mass segregation are much younger than the relaxation time ($T_{\rm E}$), which indicates that the dynamic evolution of clusters is not enough to cause the mass segregation \citep{2007AJ....134.1368C}, so the observed mass segregation may result from star formation process or the initial state of giant molecular clouds \citep{2007MNRAS.381L..40D,2010MNRAS.405..401D}. Hence, the formation mechanism of mass segregation in star clusters is still controversial \citep{2018MNRAS.473..849D,2018A&A...615A...9P} and requires further research on more clusters.

\begin{figure}[ht!]
\centering
%\plotone{m_seg}
\includegraphics[width=0.5\textwidth, angle=0]{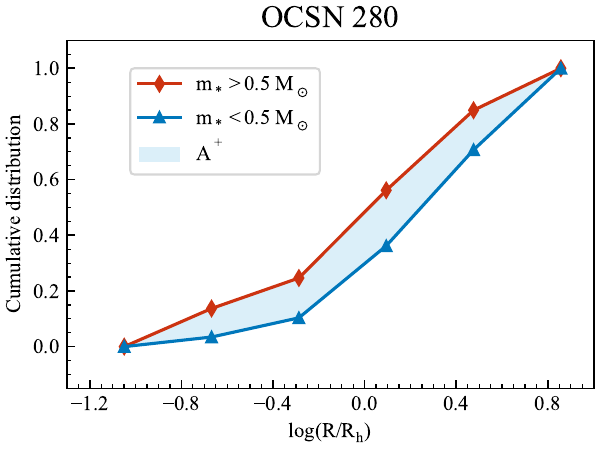}
\caption{Cumulative radial number distribution of OCSN 280. Red diamonds and blue triangles represent the radial number distribution of high- and low-mass stars, respectively. The X-axis is a logarithmic radius normalized using the half-mass radius. The index $A^+$ is calculated from the area difference between high- and low-mass samples, which can be quantified as the degree of mass segregation. 
\label{fig:m_seg}}
\end{figure}

To investigate the mass segregation in our sample, we use $0.5\,\rm M_\odot$ as the split point to divide stars into massive and low-mass parts and calculate the cumulative number of stars as a function of $\log(R/R_h)$ respectively, where $R$ is the radial distance from the cluster center and the $R_{\rm h}$ is the half-mass radius of the cluster. Furthermore, we employed the index $A^+$ to quantify the degree of mass segregation for each cluster, which is defined as the difference in the area between stars samples with different masses on the cumulative radial distribution diagram \citep{2016ApJ...833..252A}. For a star cluster without mass segregation, the spatial distribution of its massive and low-mass stars is consistent, and $A^+$ tends to 0. If mass segregation occurs within the star cluster, the massive stars will concentrate on the cluster's central region. They will occupy a larger area on the cumulative radial distribution diagram compared to the low-mass stars, resulting in a positive value for $A^+$. We have presented Figure~\ref{fig:m_seg} to illustrate the cumulative radial distribution of the cluster OCSN 280. This implies that the higher value of $A^+$ (sky-blue area) corresponds to greater mass segregation.

\begin{figure}[ht!]
\centering
%\plotone{A}
\includegraphics[width=0.38\textwidth, angle=0]{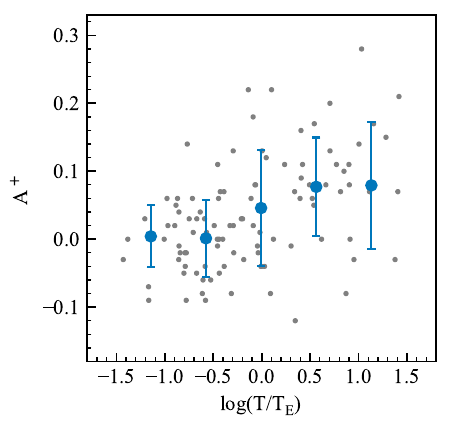}
\caption{ Distribution of mass segregation parameter $A^+$ as a function of dynamic evolution timescale. The blue dots and error bars represent each bin's mean value and corresponding standard deviation. Because older star clusters undergo more dynamic evolution processes, their degree of mass segregation increases with the cluster age.
\label{fig:A}}
\end{figure}

Figure~\ref{fig:A} displays the distribution of mass segregation parameter $A^+$ as a function of dynamic evolution timescale $\log(T/T_E)$. For each star cluster, we adopted the ratio of the cluster age to the relaxation timescale to represent the dynamic state of the cluster, while the relaxation timescale $T_{\rm E}$ is calculated by the formula given by \citet{2020MNRAS.495.2496M}:

\begin{equation} 
    T_E = \frac{8.9\times10^5\times(\frac{NR_h^3}{\bar{m}})^{0.5}}{\log(0.4N)}
    \label{equ:TE}
\end{equation}

where $N$ is the number of member stars, and $R_{\rm h}$ is the cluster's half-mass radius (pc). $\bar m$ is the average mass of cluster members ($\rm M_\odot$). Our sample demonstrates that the degree of mass segregation rises with the cluster age, indicating that the dynamic evolution of the cluster mainly causes the mass segregation. 

Furthermore, in order to investigate mass segregation during the initial phase of star cluster formation \citep{2007ApJ...655L..45M,2019MNRAS.488.1635A}, we selected 28 dynamically young clusters (corresponding to the top 25\% of dynamic age) to verify whether mass segregation had occurred. Ultimately, we discovered that 12 clusters demonstrated a positive $A^+$. This suggests that primordial mass segregation exists in clusters, although not universal. 

\subsection{Binary} \label{subsec:bin}

Binary stars are pretty common in stellar systems. A thorough comprehension of binary stars is essential to ascertain the fundamental properties of a star cluster, such as the mass function and total mass. The binary fraction and $q$ distribution are utilized to characterize the statistical properties of binary stars within the stellar cluster. 

\begin{figure}[ht!]
\centering
%\plotone{BF_dist}
\includegraphics[width=0.38\textwidth, angle=0]{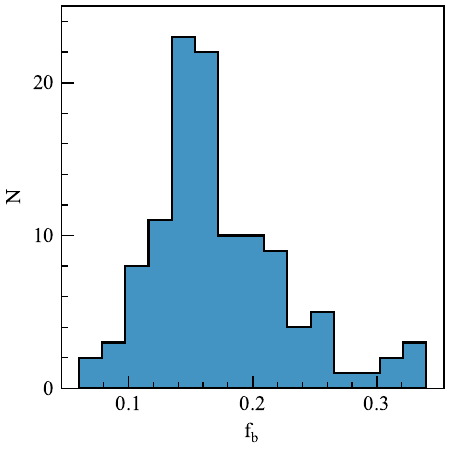}
\caption{Histogram of the binary fraction for 114 OCs. The median value of the binary fraction is 17\%.}
\label{fig:BF_dist}
\end{figure}

Figure~\ref{fig:BF_dist} shows the binary fraction distribution of 114 OCs, with a range from 6\% to 34\% and a mean value of 17.5\%. The 15th, 50th, and 85th percentiles of binary fraction are 12\%, 17\%, and 22\%, respectively. In contrast to the binary fraction of approximately 30\% reported in previous studies of OCs, \citep{2013MNRAS.434.3236K,2016MNRAS.457.1028S,2019ApJ...874..127B,2020ApJ...901...49L}, the binary fraction in this work appears to be relatively lower. This is because, to avoid counting single stars with large photometric errors as binary stars, we excluded the binary stars whose $q$ is less than 0.5 ($q<0.5$), which means the binary fraction in this work only represents the binary fraction with high $q$. As a comparison, the multiplicity fraction of resolved systems with $q>0.6$ is 15\% in \citet{2023arXiv230111061D}. In addition, we note that \citet{2020ApJ...903...93N} obtained the binary fraction of 12 OCs based on the fitting of the synthetic CMD, while the $q$ covered all binaries in the range of $0-1$. There are four common clusters (OCSN 142, OCSN 219, OCSN 228, and OCSN 259) between our sample and their sample, with binary fractions of [9\%, 13\%, 12\%, 10\%] and [18\%, 21\%, 24\%, 28\%], respectively. Given that we only calculated the binary fraction of $q \geq 0.5$ in our sample, and the proportion of these binary stars happens to be half of all binaries (the comparing result with \citet{2020ApJ...903...93N}), this may also indicate that the number of high-$q$ binaries and low-$q$ binaries is similar.

\subsection{Comparison with Other Works} \label{sec:dis}
To verify the reliability and further analyze our results, we employed the clusters in common between our catalog and other literature catalogs that provide the mass function and the binary fraction. All relevant tables are listed in the Appendix B for reference.

\begin{figure}[ht!]
\centering
\includegraphics[width=0.4\textwidth, angle=0]{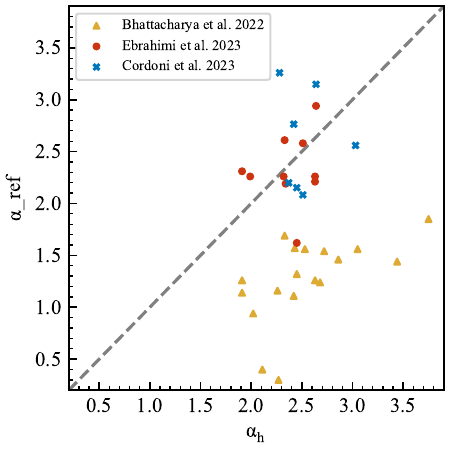}
\caption{A comparison of PDMF between this work ($\alpha_{\rm h}$) and previous studies ($\alpha$\_ref) referenced from \citet{2022MNRAS.516.5637E}, \citet{Cordoni_2023}, and \citet{2022MNRAS.517.3525B}, which include 10, 8, and 18 common OCs, respectively.}
\label{fig:alpha_com}
\end{figure}

As we described in Sect.~\ref{sec:intro}, many studies have reported their measurement results of the OCs mass function using the {\it Gaia} data. 
Cross-matching with the literature catalogs, there are 8, 10, and 18 common OCs obtained from  \citet{Cordoni_2023}, \citet{2022MNRAS.516.5637E} and \citet{2022MNRAS.517.3525B}, respectively. Considering that the mass function slope ($\alpha$\_ref) for some common OCs is fitted by a single power-law in the entire mass range, in Figure~\ref{fig:alpha_com}, we only compared the slope of the high-mass end in this work with other literature results. It is noted that only the $\alpha$\_ref from \citet{2022MNRAS.517.3525B} is underestimated compared to our result. This is because these OCs are older, with a mean $\log(t)$ of 8.20, and fitted with a single power-law. Due to mass segregation and dynamic evolution, many low-mass stars in old-age OCs will be stripped out (see Sect~\ref{subsec:PDMF}), which leads to the $\alpha$ of an entire mass range being flatter than in the high-mass end. In other words, the $\alpha$ values measured with different mass ranges, especially for old-age OCs, will present significant discrepancies. 

\begin{figure}[ht!]
\centering
\includegraphics[width=0.43\textwidth, angle=0]{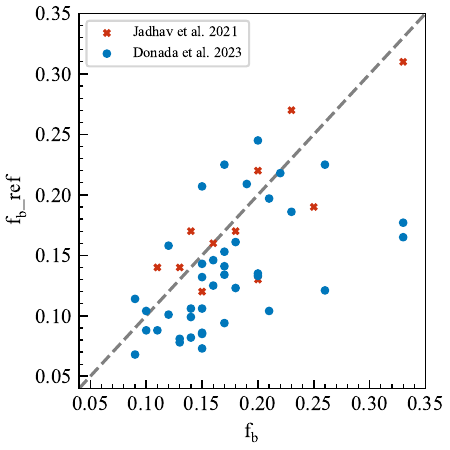}
\caption{A comparison of binary fraction between this work ($f_b$) and previous works ($f_b$\_ref) referenced from \citet{2023arXiv230111061D} and \citet{2021AJ....162..264J}, which include 11 and 40 common OCs.}
\label{fig:fb_com}
\end{figure}

For the comparison of binary fraction results, we cross-matched with the catalogs of \citet{2021AJ....162..264J} and \citet{2023arXiv230111061D} and found  11, 40 common OCs, respectively. Figure~\ref{fig:fb_com} shows the comparison result. Generally, the binary fractions in the two catalogs are consistent with our results. The mean binary fraction for  \citet{2021AJ....162..264J}, \citet{2023arXiv230111061D}, and this work are 18\%, 14\%, 17\%, respectively. Furthermore, since the mass ratio used for binary determination is 0.6 in literature, the binary fraction in this work ($q \geq 0.5$) is slightly higher than others.

\section{Summary} \label{sec:sum}

Using the catalog of nearby OCs provided by \citetalias{2023ApJS..265...12Q}, we estimated the stellar mass of 35736 member stars in 114 OCs. Unlike the traditional isochrone matching method, we utilized the RL matching approach, which provides a better correlation with the observed CMD to determine the stellar mass of each member star and the $q$ of each binary star. By adopting each cluster's stellar masses, we detailed the mass function, binary fraction, and mass segregation of clusters in our sample.

The main results of this work are as follows:
\begin{enumerate}
    \item[(1)] To improve the shortcomings in fitting CMD by the Padova isochrone at the low-mass end, we adopted the RL matching to estimate the stellar mass of members. Our results indicate that the RL matching approach provided the stellar masses with higher precision and derived the $q$ of binary stars more reliably, especially for the low-mass stars.   
    \item[(2)] We employed a two-part segmented linear function to fit the PDMF. The median values of $\alpha$ at the high- and low-mass end are 2.65 and 0.95, respectively. The $\alpha$ value is roughly consistent with the multiple-part power-law IMF in \citet{2001MNRAS.322..231K}. However, a detailed analysis reveals that the IMF of low-mass stars is more likely to exhibit a log-normal pattern after excluding older star clusters that have undergone significant dynamic evolution.
    \item[(3)] We calculated the half-mass radius $R_{\rm h}$ and the relaxation time $T_{\rm E}$ for each cluster and adopted the index $A^+$ to quantify the degree of mass segregation. Since older clusters suffered a more dynamic evolution process, the degree of mass segregation rises with cluster age. In addition, we also found primordial mass segregation in young clusters, but this phenomenon is not ubiquitous. 
    \item[(4)] We estimated the binary fraction with $q \geq 0.5$ for 114 OCs, varying from 6\% to 34\% with a median of 17\%. Compared to the 30\% binary fraction in other literature, the smaller value in this work may be due to only counting the binary stars with high $q$. On the other hand, according to the binary fraction results with the $q$ of $0-1$, the binary fraction in this work may indicate that the number of high-$q$ binaries and low-$q$ binaries is similar.
\end{enumerate}

The high accuracy of {\it Gaia} data significantly improves the membership determination of OCs. With the help of more reliable member star identification results, more and more evidence demonstrates that OCs not only have a centralized core but also present a more extended halo or filamentary substructure in the outskirts \citep{2019A&A...621L...3M,2019A&A...624A..34Z,2019A&A...621L...2R,2020ApJ...889...99Z,2022RAA....22e5022B}. By further research, it is shown that OCs exhibit a double-component structural feature, consisting of dense core components (Kings profile) and extended halo components (logarithmic Gaussian profile) \citep{2022AJ....164...54Z}. It can be expected that further statistical research on the PDMF of different components in OCs will provide new insight into the comprehensive understanding of the formation and evolution of OCs. 

\section*{Acknowledgments}
We thank the anonymous referee for the instructive comments and suggestions, which helped us immensely improve the paper. This work is supported by the National Natural Science Foundation of China (NSFC) through grants 12090040, 12090042, and 12073060. 
Jing Zhong would like to acknowledge the National Key R\&D Program of China No. 2019YFA0405501, the Youth Innovation Promotion Association CAS, and the Shanghai Science and Technology Program (20590760800).
Li Chen acknowledges the science research grants from the China Manned Space Project with NO. CMS-CSST-2021-A08.
Songmei Qin acknowledges the financial support provided by the China Scholarship Council program (Grant No. 202304910547).

This work has made use of data from the European Space Agency (ESA) mission {\it Gaia} (\url{https://www.cosmos.esa.int/gaia}), processed by the {\it Gaia} Data Processing and Analysis Consortium (DPAC,\url{https://www.cosmos.esa.int/web/gaia/dpac/consortium}). Funding for the DPAC has been provided by national institutions, in particular, the institutions participating in the {\it Gaia} Multilateral Agreement.

\section*{Appendix A} \label{appa}
We provide a catalog to present the fundamental properties of 114 nearby OCs, while the description of parameters is listed in Table~\ref{tab:cat}. Columns $2-5$ list the CMD fitting parameters $\log(t)$, $E(B-V)$, $DM$, and $Z$, obtained by the isochrone fitting with RLSSS. Columns $6-8$ list the statistical parameters, including the mass of cluster $M_{\rm tot}$, the number of member stars $N$, and the binary fraction $f_{\rm b}$ with $q \geq 0.5$. It is noted that the $M_{\rm tot}$ and N also include the contributions of secondary stars in binary systems. Columns $9-14$ are the derived parameters of the mass function. Columns $15-17$ list other useful parameters, including the half-mass radius $R_{\rm h}$, the relaxation time $T_{\rm E}$, and the mass segregation index $A^+$.

\begin{table*}[h!]
\centering
\caption{Description of the catalog of cluster properties.}
\label{tab:cat}
\begin{tabular}{l c l}
\hline
Column &  Unit   & Description \\
\hline
Cluster & - & Cluster name in \citetalias{2023ApJS..265...12Q} \\
$\log(t)$ & - & Cluster age determined by RL fitting with isochrone  \\
$E(B-V)$ & mag & Cluster reddening determined by RL fitting with isochrone  \\
$DM$ & mag & Cluster distance modulus determined by RL fitting with isochrone  \\
$Z$ & dex & Cluster metallicity determined by RL fitting with isochrone  \\
$M_{\rm tot}$ & $M_\odot$ & The total mass of cluster  \\
$N$ & - & The number of cluster members  \\
$f_{\rm b}$ & - & The binary fraction of cluster  \\
$m_{\rm c}$ & $M_\odot$ & The mass of segmentation point  \\
$\rm MR_{h}$ & $M_\odot$ & The mass range used to fit $\alpha_{\rm h}$  \\
$\alpha_{\rm h}$ & - & The index $\alpha$ of PDMF at the high-mass region  \\
${\rm err}\_\alpha_{\rm h}$ & - & The measured uncertainty of $\alpha_{\rm h}$  \\
$\rm MR_{l}$ & $M_\odot$ & The mass range used to fit $\alpha_{\rm l}$  \\
$\alpha_{\rm l}$ & - & The index $\alpha$ of PDMF at the low-mass region  \\
${\rm err}\_\alpha_{\rm l}$ & - & The measured uncertainty of $\alpha_{\rm l}$  \\
$R_{\rm h}$ & pc & The half-mass-radius of cluster \\
$T_{\rm E}$ & Myr & The relaxation time of cluster \\
$A^+$ & - & The index of mass segregation  \\
\hline
\multicolumn{3}{l}{{\sc Notes:}  This table is available in its entirety in machine-readable form.}\\
\end{tabular}
\end{table*}

\section*{Appendix B} \label{appb}
Table~\ref{tab:com_m} compares the mass function of common OCs between our result and the three references. Columns $1-4$ provide information on the cluster name (from \citetalias{2023ApJS..265...12Q}), age, the mass range used to fit the mass function, and the mass function slope at the high-mass region in this study, respectively. Columns $5-8$ display the cluster name (from the corresponding literature), the mass range used to fit the mass function, the mass function slope, and the corresponding reference, respectively.

In Table~\ref{tab:com_f}, we compare the binary fraction of common OCs between our result and the literature results. Columns $1-2$ present this work's cluster name (from \citetalias{2023ApJS..265...12Q}) and the derived binary fraction. Columns $3-5$ list the cluster name (from the corresponding literature), binary fraction, and the corresponding reference.

\begin{table*}
  \centering
  \caption{Comparing the mass function of common OCs with other works.}
  \label{tab:com_m}
  \begin{tabular}{c c c c | c c c c}
    \toprule
    Name & $\log(t)$ & $\rm MR_{h}$ & $\alpha_{\rm h}$ & Name\_ref & MR\_ref & $\alpha$\_ref & Reference \\
    \midrule
    OCSN 130 & 8.150 & $0.66-1.86$ & $2.26 \pm 0.38$ & ASCC 41 & {} & $1.16 \pm 0.05$ & \citet{2022MNRAS.517.3525B} \\
    OCSN 138 & 8.250 & $0.84-4.01$ & $2.53 \pm 0.55$ & Alessi 5 & {} & $1.56 \pm 0.24$ & \citet{2022MNRAS.517.3525B} \\
    OCSN 142 & 7.925 & $1.23-5.43$ & $3.75 \pm 0.42$ & Alessi 24 & {} & $1.85 \pm 0.14$ & \citet{2022MNRAS.517.3525B} \\
    OCSN 147 & 7.750 & $0.55-6.08$ & $2.86 \pm 0.16$ & BH 99 & {} & $1.46 \pm 0.26$ & \citet{2022MNRAS.517.3525B} \\
    OCSN 148 & 7.775 & $0.44-6.11$ & $2.72 \pm 0.17$ & BH 164 & {} & $1.54 \pm 0.10$ & \citet{2022MNRAS.517.3525B} \\
    OCSN 204 & 7.875 & $0.84-4.58$ & $2.63 \pm 0.57$ & IC 2391 & $0.2-6.0$ & $2.21 \pm 0.13$ & \citet{2022MNRAS.516.5637E} \\
    OCSN 205 & 7.800 & $0.18-4.87$ & $1.91 \pm 0.09$ & IC 2602 & {} & $1.26 \pm 0.10$ & \citet{2022MNRAS.517.3525B} \\
    {} & {} & {} & {} & IC 2602 & $0.2-6.5$ & $2.31 \pm 0.21$ & \citet{2022MNRAS.516.5637E} \\
    OCSN 206 & 8.350 & $0.95-3.46$ & $2.45 \pm 0.50$ & IC 4665 & {} & $1.32 \pm 0.03$ & \citet{2022MNRAS.517.3525B} \\
    OCSN 207 & 9.250 & $0.77-1.59$ & $2.27 \pm 0.50$ & IC 4756 & {} & $0.30 \pm 0.23$ & \citet{2022MNRAS.517.3525B} \\
    OCSN 218 & 8.200 & $0.35-4.11$ & $2.32 \pm 0.12$ & $\alpha$-Per & $0.2-5.6$ & $2.26 \pm 0.15$ & \citet{2022MNRAS.516.5637E} \\
    OCSN 219 & 8.475 & $0.28-3.19$ & $2.33 \pm 0.11$ & Pleiades & {} & $1.68 \pm 0.18$ & \citet{2022MNRAS.517.3525B} \\
    {} & {} & {} & {} & Pleiades & $0.2-4.9$ & $2.61 \pm 0.14$ & \citet{2022MNRAS.516.5637E} \\
    {} & {} & {} & {} & Melotte 22 & $>1$ & $0.0 \pm 0.0$ & \citet{Cordoni_2023} \\
    OCSN 220 & 8.800 & $0.72-2.12$ & $2.45 \pm 0.49$ & Hyades & $0.2-2.5$ & $1.62 \pm 0.18$ & \citet{2022MNRAS.516.5637E} \\
    OCSN 222 & 8.525 & $0.69-3.05$ & $3.05 \pm 0.19$ & NGC 1039 & {} & $1.56 \pm 0.22$ & \citet{2022MNRAS.517.3525B} \\
    OCSN 228 & 8.800 & $0.58-2.42$ & $3.03 \pm 0.23$ & NGC 2281 & $>1$ & $2.559 \pm 0.0$ & \citet{Cordoni_2023} \\
    OCSN 229 & 8.100 & $0.58-4.61$ & $2.48 \pm 0.10$ & NGC 2422 & $>1$ & $3.259 \pm 0.023$ & \citet{Cordoni_2023} \\
    OCSN 230 & 7.825 & $0.36-5.28$ & $2.34 \pm 0.27$ & NGC 2451A & $0.2-6.1$ & $2.19 \pm 0.17$ & \citet{2022MNRAS.516.5637E} \\
    OCSN 231 & 7.825 & $0.45-6.03$ & $2.43 \pm 0.20$ & NGC 2451B & {} & $1.57 \pm 0.33$ & \citet{2022MNRAS.517.3525B} \\
    OCSN 232 & 8.225 & $0.48-3.98$ & $2.64 \pm 0.10$ & NGC 2516 & $0.3-3.3$ & $2.94 \pm 0.21$ & \citet{2022MNRAS.516.5637E} \\
    {} & {} & {} & {} & NGC 2516 & $>1$ & $3.148 \pm 0.0$ & \citet{Cordoni_2023} \\
    OCSN 233 & 7.650 & $0.78-4.47$ & $2.51 \pm 0.54$ & NGC 2547 & $0.2-5.1$ & $2.58 \pm 0.24$ & \citet{2022MNRAS.516.5637E} \\
    {} & {} & {} & {} & NGC 2547 & $>1$ & $2.083 \pm 0.0$ & \citet{Cordoni_2023} \\
    OCSN 234 & 8.850 & $0.50-2.44$ & $2.45 \pm 0.31$ & NGC 2632 & $>1$ & $2.152 \pm 0.0$ & \citet{Cordoni_2023} \\
    OCSN 235 & 7.575 & $0.48-4.40$ & $2.02 \pm 0.22$ & NGC 3228 & {} & $0.94 \pm 0.31$ & \citet{2022MNRAS.517.3525B} \\
    OCSN 236 & 8.600 & $0.71-3.00$ & $2.37 \pm 0.16$ & NGC 3532 & $>1$ & $2.199 \pm 0.084$ & \citet{Cordoni_2023} \\
    OCSN 240 & 8.275 & $0.82-3.97$ & $2.63 \pm 0.23$ & NGC 6475 & {} & $1.26 \pm 0.25$ & \citet{2022MNRAS.517.3525B} \\
    {} & {} & {} & {} & NGC 6475 & $0.2-3.1$ & $2.26 \pm 0.22$ & \citet{2022MNRAS.516.5637E} \\
    OCSN 243 & 8.650 & $0.28-2.85$ & $1.99 \pm 0.16$ & NGC 7092 & $0.2-3.1$ & $2.26 \pm 0.29$ & \citet{2022MNRAS.516.5637E} \\
    OCSN 259 & 8.375 & $0.48-2.83$ & $1.91 \pm 0.21$ & Roslund 6 & {} & $1.14 \pm 0.20$ & \citet{2022MNRAS.517.3525B} \\
    OCSN 264 & 8.825 & $0.75-2.51$ & $2.11 \pm 0.48$ & Stock 1 & {} & $0.4 \pm 0.06$ & \citet{2022MNRAS.517.3525B} \\
    OCSN 267 & 8.300 & $0.70-3.59$ & $2.42 \pm 0.12$ & Stock 12 & {} & $1.11 \pm 0.18$ & \citet{2022MNRAS.517.3525B} \\
    {} & {} & {} & {} & Stock 12 & $>1$ & $2.764 \pm 0.054$ & \citet{Cordoni_2023} \\
    OCSN 275 & 8.325 & $0.89-3.34$ & $3.44 \pm 0.33$ & Teutsch 35 & {} & $1.44 \pm 0.38$ & \citet{2022MNRAS.517.3525B} \\
    OCSN 276 & 7.775 & $0.57-5.73$ & $2.68 \pm 0.19$ & Trumpler 10 & {} & $1.24 \pm 0.25$ & \citet{2022MNRAS.517.3525B} \\
    \bottomrule
  \end{tabular}
\end{table*}

\begin{table*}[ht]
  \centering
  \caption{Comparing the binary fraction of common OCs with other works.}
  \label{tab:com_f}
  \begin{tabular}{c c | c c c}
    \toprule
    Name & $f_b$ & Name\_ref  & $f_b$\_ref & Reference \\
    \midrule
    OCSN 127 & 0.17 & ASCC 16 & $0.153 \pm 0.033$ & \citet{2023arXiv230111061D} \\
    OCSN 128 & 0.10 & ASCC 19 & $0.088 \pm 0.035$ & \citet{2023arXiv230111061D} \\
    OCSN 129 & 0.15 & ASCC 21 & $0.207 \pm 0.116$ & \citet{2023arXiv230111061D} \\
    OCSN 130 & 0.15 & ASCC 41 & $0.143 \pm 0.035$ & \citet{2023arXiv230111061D} \\
    OCSN 131 & 0.15 & ASCC 58 & $0.073 \pm 0.020$ & \citet{2023arXiv230111061D} \\
    OCSN 134 & 0.21 & ASCC 124 & $0.197 \pm 0.087$ & \citet{2023arXiv230111061D} \\
    OCSN 135 & 0.16 & ASCC 127 & $0.146 \pm 0.038$ & \citet{2023arXiv230111061D} \\
    OCSN 137 & 0.19 & Alessi 3 & $0.209 \pm 0.044$ & \citet{2023arXiv230111061D} \\
    OCSN 139 & 0.26 & Alessi 9 & $0.121 \pm 0.033$ & \citet{2023arXiv230111061D} \\
    OCSN 142 & 0.09 & Alessi 24 & $0.068 \pm 0.022$ & \citet{2023arXiv230111061D} \\
    OCSN 146 & 0.17 & BH 23 & $0.134 \pm 0.044$ & \citet{2023arXiv230111061D} \\
    OCSN 148 & 0.15 & BH 164 & $0.106 \pm 0.025$ & \citet{2023arXiv230111061D} \\
    OCSN 191 & 0.12 & Collinder 140 & $0.158 \pm 0.042$ & \citet{2023arXiv230111061D} \\
    OCSN 194 & 0.22 & UBC 17b & $0.218 \pm 0.051$ & \citet{2023arXiv230111061D} \\
    OCSN 195 & 0.26 & Gulliver 9 & $0.225 \pm 0.043$ & \citet{2023arXiv230111061D} \\
    OCSN 207 & 0.33 & IC 4756 & $0.31 \pm 0.03$ & \citet{2021AJ....162..264J} \\
    {} & {} & IC 4756 & $0.165 \pm 0.029$ & \citet{2023arXiv230111061D} \\
    OCSN 210 & 0.09 & UBC 7 & $0.114 \pm 0.043$ & \citet{2023arXiv230111061D} \\
    OCSN 217 & 0.16 & Mamajek 4 & $0.125 \pm 0.022$ & \citet{2023arXiv230111061D} \\
    OCSN 219 & 0.13 & Pleiades & $0.14 \pm 0.02$ & \citet{2021AJ....162..264J} \\
    {} & {} & Melotte 22 & $0.078 \pm 0.009$ & \citet{2023arXiv230111061D} \\
    OCSN 221 & 0.25 & NGC 752 & $0.19 \pm 0.03$ & \citet{2021AJ....162..264J} \\
    OCSN 222 & 0.14 & NGC 1039 & $0.17 \pm 0.02$ & \citet{2021AJ....162..264J} \\
    {} & {} & NGC 1039 & $0.106 \pm 0.014$ & \citet{2023arXiv230111061D} \\
    OCSN 228 & 0.12 & NGC 2281 & $0.101 \pm 0.056$ & \citet{2023arXiv230111061D} \\
    OCSN 229 & 0.11 & NGC 2422 & $0.14 \pm 0.02$ & \citet{2021AJ....162..264J} \\
    {} & {} & NGC 2422 & $0.088 \pm 0.035$ & \citet{2023arXiv230111061D} \\
    OCSN 231 & 0.14 & NGC 2451B & $0.099 \pm 0.025$ & \citet{2023arXiv230111061D} \\
    OCSN 232 & 0.16 & NGC 2516 & $0.16 \pm 0.01$ & \citet{2021AJ....162..264J} \\
    OCSN 233 & 0.20 & NGC 2547 & $0.22 \pm 0.04$ & \citet{2021AJ....162..264J} \\
    {} & {} & NGC 2547 & $0.135 \pm 0.029$ & \citet{2023arXiv230111061D} \\
    OCSN 235 & 0.17 & NGC 3228 & $0.094 \pm 0.045$ & \citet{2023arXiv230111061D} \\
    OCSN 236 & 0.20 & NGC 3532 & $0.13 \pm 0.01$ & \citet{2021AJ....162..264J} \\
    OCSN 238 & 0.23 & NGC 6281 & $0.27 \pm 0.02$ & \citet{2021AJ....162..264J} \\
    OCSN 239 & 0.18 & NGC 6405 & $0.17 \pm 0.02$ & \citet{2021AJ....162..264J} \\
    {} & {} & NGC 6405 & $0.123 \pm 0.017$ & \citet{2023arXiv230111061D} \\
    OCSN 243 & 0.21 & NGC 7092 & $0.104 \pm 0.025$ & \citet{2023arXiv230111061D} \\
    OCSN 250 & 0.15 & Platais 3 & $0.086 \pm 0.033$ & \citet{2023arXiv230111061D} \\
    OCSN 256 & 0.18 & RSG 5 & $0.161 \pm 0.033$ & \citet{2023arXiv230111061D} \\
    OCSN 261 & 0.33 & Ruprecht 147 & $0.177 \pm 0.042$ & \citet{2023arXiv230111061D} \\
    OCSN 263 & 0.17 & Stephenson 1 & $0.255 \pm 0.060$ & \citet{2023arXiv230111061D} \\
    OCSN 267 & 0.17 & Stock 12 & $0.141 \pm 0.035$ & \citet{2023arXiv230111061D} \\
    OCSN 275 & 0.13 & Teutsch 35 & $0.081 \pm 0.024$ & \citet{2023arXiv230111061D} \\
    OCSN 276 & 0.15 & Trumpler 10 & $0.12 \pm 0.02$ & \citet{2021AJ....162..264J} \\
    OCSN 278 & 0.10 & UBC 1 & $0.104 \pm 0.032$ & \citet{2023arXiv230111061D} \\
    OCSN 279 & 0.20 & UBC 8 & $0.245 \pm 0.041$ & \citet{2023arXiv230111061D} \\
    OCSN 283 & 0.14 & UBC 17a & $0.082 \pm 0.025$ & \citet{2023arXiv230111061D} \\
    OCSN 288 & 0.23 & UBC 32 & $0.186 \pm 0.041$ & \citet{2023arXiv230111061D} \\
    OCSN 312 & 0.20 & UPK 535 & $0.133 \pm 0.039$ & \citet{2023arXiv230111061D} \\
    OCSN 314 & 0.15 & UPK 545 & $0.085 \pm 0.021$ & \citet{2023arXiv230111061D} \\
    OCSN 319 & 0.15 & UPK 612 & $0.132 \pm 0.049$ & \citet{2023arXiv230111061D} \\
    \bottomrule
  \end{tabular}
\end{table*}

\bibliography{sample631}{}
\bibliographystyle{aasjournal}

\end{CJK*}
\end{document}